\documentclass[aps,twocolumn,superscriptaddress,letterpaper]{revtex4}
\usepackage{graphicx,amssymb,bm,natbib,amsfonts,amsmath}
\usepackage{hyperref}

\begin{document}

\title{Core Level and Valence Band Study of LaFeAsO$_{0.9}$F$_{0.1}$}

\author{D.R. Garcia}
\affiliation{Department of Physics, University of California,
Berkeley, CA 94720, USA}
\affiliation{Materials Sciences Division,
Lawrence Berkeley National Laboratory, Berkeley, CA 94720, USA}
\author{C. Jozwiak}
\affiliation{Department of Physics, University of California,
Berkeley, CA 94720, USA}
\author{C.G. Hwang}
\affiliation{Materials Sciences Division,
Lawrence Berkeley National Laboratory, Berkeley, CA 94720, USA}
\author{A. Fedorov}
\affiliation{Advanced Light Source,
Lawrence Berkeley National Laboratory, Berkeley, CA 94720, USA}
\author{S.M. Hanrahan}
\affiliation{Life Sciences Division,
Lawrence Berkeley National Laboratory, Berkeley, CA 94720, USA}
\author{S.D. Wilson}
\affiliation{Materials Sciences Division,
Lawrence Berkeley National Laboratory, Berkeley, CA 94720, USA}
\author{\mbox{C.R. Rotundu}}
\affiliation{Materials Sciences Division,
Lawrence Berkeley National Laboratory, Berkeley, CA 94720, USA}
\author{B.K. Freelon}
\affiliation{Materials Sciences Division,
Lawrence Berkeley National Laboratory, Berkeley, CA 94720, USA}
\author{R.J. Birgeneau}
\affiliation{Department of Physics, University of California,
Berkeley, CA 94720, USA}
\affiliation{Materials Sciences Division, Lawrence Berkeley National Laboratory, Berkeley, CA 94720, USA}
\author{E. Bourret-Courchesne}
\affiliation{Materials Sciences Division,
Lawrence Berkeley National Laboratory, Berkeley, CA 94720, USA}
\author{A. Lanzara $^\dagger$}
\affiliation{Department of Physics, University of California,
Berkeley, CA 94720, USA}
\affiliation{Materials Sciences Division, Lawrence Berkeley National Laboratory, Berkeley, CA 94720, USA}

\date{\today}

\begin{abstract}
Using angle-integrated photoemission spectroscopy we have probed the novel LaFeAsO$_{0.9}$F$_{0.1}$ superconductor over a wide range of photon energies and temperatures.  We have provided the first full characterization of the orbital character of the VB DOS and of the magnitude of the d-p hybridization energy. Finally, we have identified two characteristic temperatures: 90K where a pseudogap-like feature appears to close and 120K where a sudden change in the DOS near E$_F$ occurs.  We associate these phenomena with the SDW magnetic ordering and the structural transition seen in the parent compound, respectively.  These results suggest the important role of electron correlation, spin physics and structural distortion in the physics of Fe-based superconductors. 
\end{abstract}

\maketitle
\section{Introduction}
    With the recent discovery of remarkably high temperature superconductivity in the fluorine doped LaFeAsO$_{1-x}$F$_{x}$ \cite{KamiharafirstLaFeAsO}, there has been an enormous effort to understand these materials in order to discover whether the nature of their superconductivity is similar to that seen in the well known high temperature superconducting cuprates or if it is of a completely different type altogether.  One reason for the potential uniqueness of these compounds is that bands related to Fe appear to dominate the near E$_F$ band structure and are hence responsible for superconductivity \cite{MatheoryLDA}.  This is unlike the superconducting cuprates where superconductivity is believed to occur in the CuO plane and, thus, could open the door for additional magnetic degrees of freedom in the FeAs systems.  This situation, together with the proposed strong hybridization between the Fe 3d and As 4p orbitals needed to explain the anomalously small value of the Fe magnetic moment \cite{WuTheoryPhilips}, begins to place the Fe-based superconductors in a strong correlation regime as is already the case for the cuprate superconductors.  Additionally, the LaFeAsO parent compound has been observed to have both an anti-ferromagnetic spin density wave (SDW) phenomenon at $\sim$134K, along with a structural distortion at $\sim$150K \cite{Cruzneutronscattering, KlaussMossbauerconfirmneutron}.  This latter phenomena is believed to be an important parameter for modulating both the d-p hybridization and the onset of the antiferromagnetic fluctuation \cite{WuTheoryPhilips}.  The presence of these phenomena in the LaFeAsO$_{1-x}$F$_{x}$ phase diagram and their apparent competition with the superconducting phase further underscore the complex physics of these materials and their near E$_F$ properties \cite{Cruzneutronscattering, Dongcompetingorder, KitaoMossbauerconfirmneutron}.  

\begin{figure}
\includegraphics[width=9.0 cm]{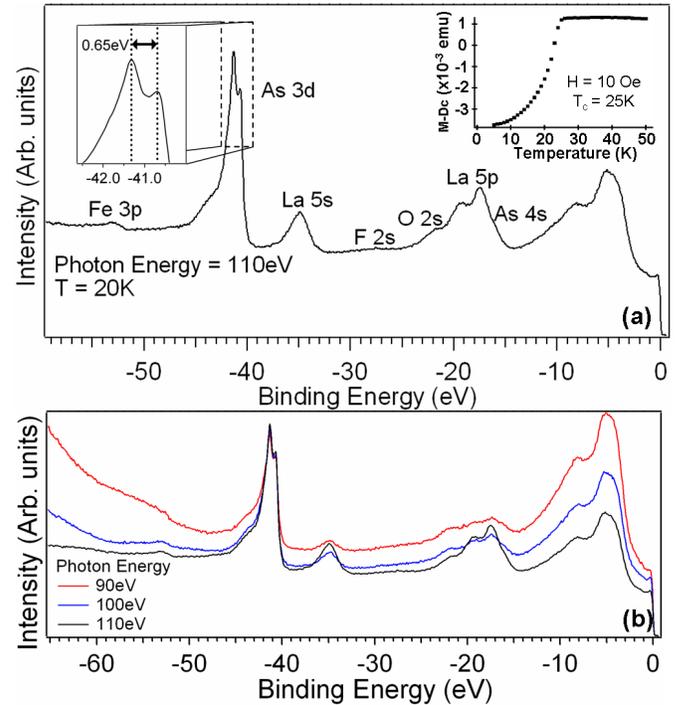} 
\caption{(color online)(a) Core level spectroscopy of LaFeAsO$_{0.9}$F$_{0.1}$ taken in the SC phase using 110eV photon energy.  The upper right inset shows the measured T$_c$ of our sample.  (b) Core level spectrum from (a) taken at three different photon energies.  The spectra are normalized to photon flux and show differing trends in intensity based on orbital character.} 
\end{figure}	

Despite the numerous angle integrated and angle resolved photoemission results that have appeared in the literature in the past few months \cite{Fengfirstpaper, ShinfirstlaserpaperLa, SatofirstLaHepaper, X.J.ZhouFirstLaserSm, X.J.Zhoucomparison, LiuNdARPES, LiuBaKFeAsARPES, ZhaoBaKFeAsARPES}, basic knowledge of the electronic structure of these materials, the importance of the correlation effects, and how the spin physics modulates the near E$_F$ states are still not fully understood.  So far, limited data on the orbital nature valence band spectra exists and no measurements of the proposed strong p-d hybridization are available.  Similarly no current understanding of the role of the structural distortion \cite{Cruzneutronscattering, KlaussMossbauerconfirmneutron} on the near E$_F$ states exist.  These concerns, along with the current controversy over the role of spin fluctuations \cite{Cruzneutronscattering, Dongcompetingorder, KitaoMossbauerconfirmneutron, MatheoryLDA, MazinimportanttheoryDFT, NowikFirstMossLa, NowikFirstMossSm, AhilanNMRLa, ZhuNernstLa, GonnelliAndreevCoexistence, DrewMuonCoexistence} in explaining the low energy features of the near E$_F$ DOS, are certainly among the fundamental aspects of these materials that must be addressed before we can come to a complete understanding of the physics of these materials.

\begin{figure}
\includegraphics[width=9.20 cm]{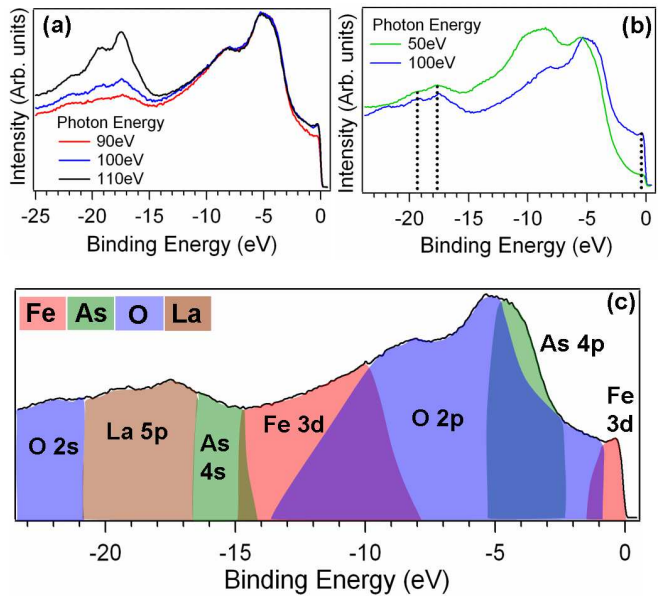}
\caption{(color online) (a) The VB region from Fig.~1b now normalized to the peak near 5eV binding energy.  This helps distinguish Fe from O orbitals in the VB.  (b) VB spectra taken with 50eV and 100eV photon energy.  The dashed lines mark the locations of the La 5p and Fe 3d peaks.  (c) A cartoon summarizing our proposed assignments of orbital character to different regions of the VB DOS based on the data from Fig.~1 and 2a-b.}  
\end{figure}
  
    In this paper, we use angle-integrated photoemission spectroscopy (AIPES) on polycrystalline LaFeAsO$_{0.9}$F$_{0.1}$ (T=25K) to address these issues.  We explore the origins of core level peaks, the valence band (VB) density of states (DOS), and near E$_F$ features over a range of photon energy and temperature.  We identify all expected core levels within our energy range and provide the first experimental characterization of the orbital character of the VB DOS.  We find an anomalously large splitting (0.65eV) of the As 3d orbital as well as an unexplained satellite feature associated with it, which allow us to provide bounds on the magnitude of the Fe/As hybridization, that are consistent with recent theoretical work, to explain the origin of the anomalously low Fe 3d magnetic moment \cite{WuTheoryPhilips}.  
Finally, we report significant changes in the near E$_F$ spectra at two characteristic temperatures, 90K and 120K, which we identify with the closing of a pseudogap-like feature and the onset of a structural transition respectively.  This work suggests that strong correlations related to spin physics, such as the phenomena observed in the undoped parent compound, need to be included in our understanding of the superconducting state in these FeAs systems.

\section{Experimental Procedure}
  Samples of polycrystalline LaFeAsO$_{0.9}$F$_{0.1}$ were synthesized by solid state reactions.  La and As metals were mixed and reacted in an evacuated sealed quartz ampoule at 500C for 12 hours, followed by a reaction at 850C for 2 hours. The formation of LaAs was confirmed by X-ray diffraction. Fe$_2$O$_3$, Fe and LaF$_3$ were then mixed with LaAs and loaded into a tantalum tube. The amount of LaF$_3$ was calculated to provide 10 percent F. The tube was crimped and sealed in an evacuated quartz ampoule. The reaction was carried out at 1150C for 50 hours. The final product: LaAsFeOF was confirmed by X-ray diffraction.  The sample has a critical temperature of 25K as determined by the diamagnetic transition in our magnetization measurements, shown in the inset of Figure 1\cite{Bobnote}.  
    Synchrotron photoemission spectroscopy data were taken at Beamline 12.0.1 at the Advanced Light Source of the Lawrence Berkeley National Laboratory using a Scienta SES100 hemispherical analyzer and also at Beamline 5.4 at the Stanford Synchrotron Radiation Laboratory, using a Scienta R4000 analyzer.  The energy resolution was 0.1eV for both the core level and VB data.  A resolution of 25meV was used for the near E$_F$ data in Figure 3, while 5meV was achieved for the data presented in Figure 4. 
Sample surfaces were prepared both by cleaving the polycrystalline rod {\em in situ}, and by a cleaving followed with gentle scraping {\em in situ} of the exposed surface using a diamond file. Both were done at a temperature less than 25K and a base pressure better than $5\times 10^{-11}$ torr.  From our experiments and the data presented, we found no quantitative differences between these two methods.

\section{Results}
\subsection{Core Level Spectra}
Figure 1 shows core level spectra taken in the superconducting phase along with the atomic orbitals associated with them. Within the energy window of Figure 1, we identify the expected La core levels: La 5s at 35.0eV, La 5p$_{1/2}$ at 19.0eV and La 5p$_{3/2}$ at 17.5eV.  Similarly, the only expected oxygen core level in this energy window is the O 2s peak, which is generally expected near 20eV in other oxides\cite{Sawatzky} and thus we identify it with the shoulder to the left of the mentioned La 5p orbitals.  A greater difficulty comes with resolving the fluorine core level peaks.  It is expected that the F 2p binding energy should be only a few eV larger than the O 2p orbital\cite{Sawatzky}.  As theory suggests and we will argue in section III.B, the spectral weight between 5 and 10 eV is dominated by the O 2p orbital, putting the likely weaker F 2p orbital contribution underneath this spectra.  Nevertheless, we interpret the weak, broad feature at 27eV as the F 2s orbital expected at a roughly 20eV higher binding energy than F 2p.  
	
	 As for the Fe core level, we can successfully resolve the Fe 3p orbital, although we are unable to resolve any potential splitting between the 3p$_{1/2}$ and 3p$_{3/2}$, most likely due to the weak intensity. Surprisingly, when we look at the As 3d core level, we observe a splitting between the 3d$_{3/2}$ (40.7eV) and 3d$_{5/2}$ (41.3eV) peaks of $\sim$0.65eV, which is much larger than what observed for As in its natural form \cite{Ascore} and, is $\sim$.05eV \cite{KoitzschLaCore} larger than observed in the undoped parent compound, LaFeAsO.  Within our resolution, the magnitude of such splitting is temperature independent, above and below T$_c$.  This large splitting suggests that we are in a regime where the hybridization between the Fe 3d and the As 4p orbitals is quite strong, as previously proposed \cite{WuTheoryPhilips, Caoabinitotheory}.  Modeling the splitting as due to hybridization between Fe 3p$_{3/2}$ and the As 3d$_{3/2}$ orbitals with a two level model \cite{private}, we can put experimental constraints on the magnitude of the Fe/As hybridization energy, finding it to be between 1.1 and 1.4eV. This is consistent with the values required to explain the anomalously small Fe magnetic moment reported for these materials  \cite{WuTheoryPhilips}.

  Finally, we did resolve two additional features in the core level study which do not directly correspond to any known orbital core level: a) a subtle shoulder at $\sim$16eV near the La 5p peaks and b) a well-defined shoulder at $\sim$44eV, to the left of the As 3d peak.  Because their binding energies are photon energy independent, neither of these features can be associated with a secondary peak or an Auger peak.  
  For the former feature, one possibility is to associate it with the currently unresolved As 4s peak, although shifted by $\sim$4eV from its expected value.  A second possibility is to consider it a satellite feature associated with the La 5p orbitals.  Photon energy dependence, as explained later on, leads us to the first conclusion, associating the feature with the weaker As 4s state, yet the reason for the apparent energy shift in this peak remains unclear.

For the latter feature, the absence of any expected additional peak within the energy range leads us to suspect it to be a satellite feature of the As 3d peaks related to the critical FeAs layer, particularly to the Fe 3p-As 3d strong hybridization previously proposed to explain the observed As 3d splitting.  This feature appears to be unchanged above and below T$_c$.  Modeling the spectra near the As 3d peaks with multiple Lorenzian peaks, we estimate the peak in this satellite feature to appear around $\sim$43eV, a separation of more than 1.5eV from the As 3d$_{3/2}$ peak. Satellite structures in fact, arise from Coulombic attraction between a core hole, created by photoemission and the VB states and have been instrumental in the study of other materials: the Cu 2p orbital in high T$_c$ cuprates\cite{CupShen} and the Ru 3d orbital in the ruthenates\cite{RuthKim}.  One possiblity is that the considerable As 4p character that theory expects around 1.5eV above E$_{F} $\cite{MatheoryLDA, Caoabinitotheory, Haulefirste-structure, SinghTheory} could cause a satellite peak to form around 2-3 eV from the As 3d peaks as the As 3d core electrons are photoemitted, potentially explaining the observed shoulder.

To further confirm our analysis, we have performed a detailed photon energy dependent study of the entire core level region.  Photon energy dependent photoemission has the advantage that the photoemission cross section changes uniquely for different elements.  Core level spectra at a few characteristic photon energies are shown in Fig.~1b.  A clear enhancement of the peaks near 35, 19, and 18 eV, already identified with the La 5s and the 5p orbitals, is observed when the photon energy approaches the La 4d core level resonance (at $\sim$105eV).  This supports the La nature of these peaks and gives weight to the prior hypothesis that the additional energy shoulder near 16eV, which is not similarly enhanced but appears relatively stronger in the 90eV curve, is indeed unrelated to La and is of an As 4s origin. 

\begin{figure}
\includegraphics[width=8.100 cm]{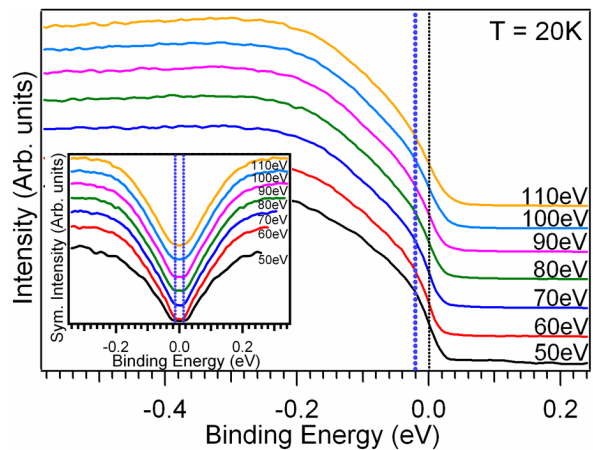}
\caption{(color online) Near E$_F$ data taken over a range of photon energies from 50 - 110eV at T=20K.  The blue dotted line indicates the energy scale of the spectral feature 15meV below E$_F$. The inset symmetrizes the data to remove contributions from the Fermi function.  The blue dotted lines indicate the $2\Delta\sim$30meV gap feature which we see at all energies.} 
\end{figure}	

\subsection{Valence Band Spectra}
		Having identified all the core level peaks in the spectra, the next crucial step is to identify the nature of the valence band region below 15 eV.  Although much theoretical work has proposed that this region is mostly dominated by Oxygen and Iron contributions \cite{MatheoryLDA,Haulefirste-structure,SinghTheory,Caoabinitotheory}, so far there has not been a complete experimental characterization of this region.  Photon energy dependent photoemissison can provide direct information on the DOS since, as mentioned, the photoemission cross section changes uniquely for different elements.  In order to better discern the relative O and Fe concentrations, Fig.~2a focuses on the VB region between E$_F$ and 25eV in binding energy.  The spectra are renormalized to the peak near 5eV, which allows one to distinguish the 3-15eV VB region from the sharp feature seen at $\sim$0.3eV.  Since the photoemission cross-section is known to decrease faster for O than Fe over this photon energy range \cite{Cross-section}, seeing the increase in the 0.3eV feature relative to the remaining VB region helps confirm its orbital character as Fe 3d while the remaining VB has a majority O 2p character, a fact also hinted to by prior DOS calculations \cite{Haulefirste-structure, SinghTheory, MatheoryLDA}. We also note that although we clearly observe this Fe 3d peak at nearly 0.3eV when using higher photon energies (50-110eV), the peak does appear to be shifted closer to 0.2eV when we employ a lower photon energy (23eV).  This is not surprising since the peak is quite likely composed of multiple bands which may also be relatively enhanced with different photon energies.

\begin{figure*}
\includegraphics[width=18 cm]{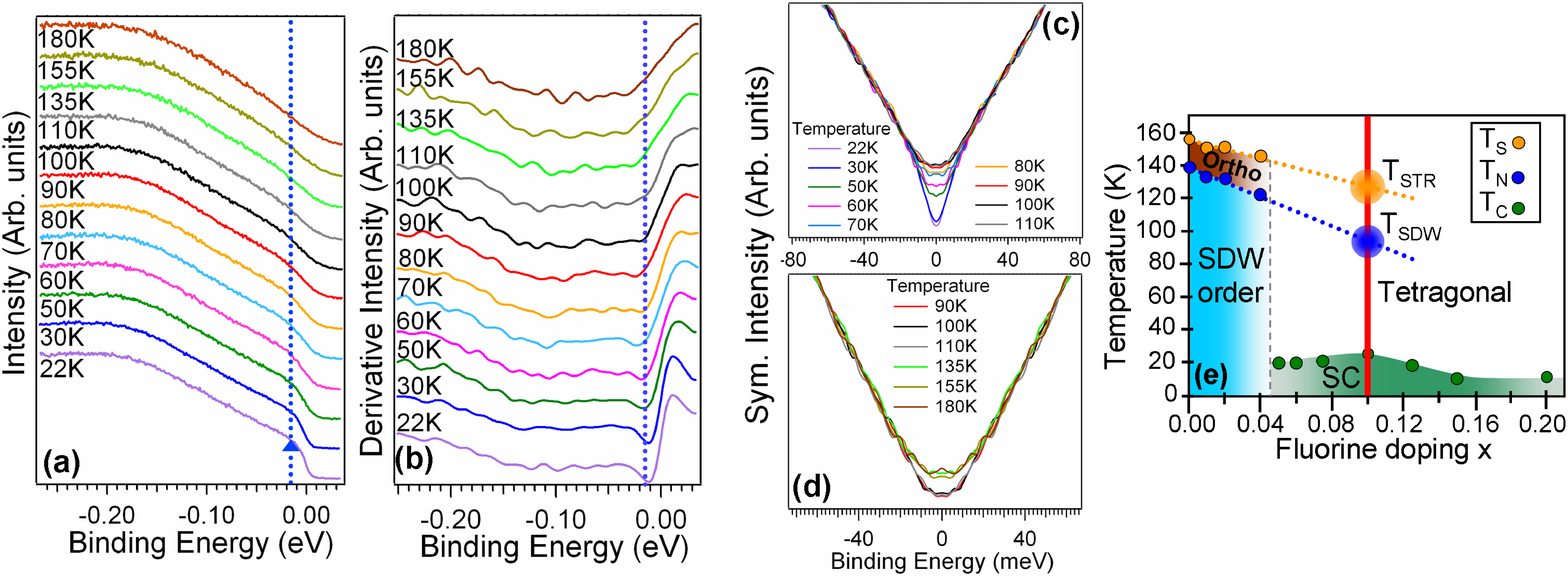}
\caption{(color online) (a) Higher resolution temperature dependent data taken with 23eV photon energy. The blue dotted line with arrow indicates the near E$_F$ features noted in Figure 3 at 15meV. (b) First derivative taken of the data from (a) symmetrized. The temperature dependent dip near E$_F$ corresponds to the 15meV feature and this energy scale is, again, marked by the blue dotted line. (c-d) Symmetrized data from (a) near E$_F$.  In (c), we see the gradual temperature evolution of the 15meV pseudogap feature until the evolution stops around 90K.  In (d), the temperature increase continues with a second, abrupt change in the near E$_F$ DOS lineshape between 110 and 135K.  (e) Using phase diagram data from \cite{LuetkensPhaseLa}, we take the T$_N$ temperatures (blue dots) which define the SDW phase and extrapolate them (blue line) to determine the potential transition temperature for the SDW phase at doping x=0.1 (red line), finding a T$_{SDW}\sim$90K, associating this with (c).  This is similarly done for the orthorhombic-tetragonal structure transition temperatures, T$_S$ (yellow dots), to find the potential transition temperature at x=0.1, T$_{STR}\sim$120K, associating this with (d).}   
\end{figure*}


	   To explore additional orbital character within the VB, Fig.~2b compares spectra taken with both 100eV and 50eV light.  Although the proposed La 5p and Fe 3d peaks only change in relative peak intensity, the remaining VB is greatly affected, particularly near 10eV with new peak features being enhanced.  The proximity of the 50eV data to the Fe 3p resonance at 53eV leads us to focus on two points.
	     First, the sudden appearance of a feature near 13eV in the 50eV data suggests an Fe orbital character to this region within the VB.  Indeed, a closer examination of that energy region (10-14eV) in Fig.~2a shows that the spectral intensity actually does increase similarly to the increase seen in the 0.3eV feature already attributed to Fe 3d.  Thus, one can speculate that this region of the VB DOS has a larger Fe 3d contribution than at slightly lower binding energy.  That would explain the origin of the additional peak structure at $\sim$10eV appearing in the 50eV data when compared to the higher photon energy data.  
	   Second, in light of this analysis, we can also attempt to make some sense of the strong $\sim$5eV feature in the DOS.  A close examination of the feature at higher photon energy clearly shows that at least two peaks are likely responsible for the slightly asymmetric lineshape.  The change in lineshape between the 100eV and 50eV data leads us to suspect that an additional, non-oxygen orbital may be contributing to the overall spectral weight.  From our analysis of Fig.~1b, La seems unlikely.  Furthermore, the bandwidth change between 100eV and 50eV of this feature appears to become narrower, suggesting the additional spectral weight is weakened at 50eV, making Fe 3d unlikely.  In light of these observations, we propose that although the primary character of this peak appears to be O 2p, the additional spectral weight at lower binding energy is related to As, most likely the previously unaccounted for 4p orbital whose hybridization with Fe 3d was discussed earlier.  Fig.~2c provides a cartoon to qualitatively summarize our conclusions regarding the orbital character of the VB DOS.

\subsection{Near-E$_F$ Spectra}	   
	 We now turn our attention from the VB to the near E$_F$ spectra and its evolution with photon energy and temperature.  Although AIPES measurements have already been carried out in these materials \cite{Fengfirstpaper,ShinfirstlaserpaperLa, SatofirstLaHepaper, X.J.Zhoucomparison}, and recent ARPES works focusing on the near E$_F$ region of related FeAs systems are now becoming available \cite{LiuNdARPES, LiuBaKFeAsARPES, ZhaoBaKFeAsARPES}, the current studies are lacking a detailed temperature dependence analysis, thus leading to controversy on the origin of the different energy scales identified \cite{ShinfirstlaserpaperLa, SatofirstLaHepaper, X.J.ZhouFirstLaserSm, X.J.Zhoucomparison}.  With AIPES being able to directly measure the DOS as we have done for the VB region, we can focus near E$_F$ to discern gap and pseudogap features in the spectra \cite{YoshidaAIPES}.  
	 
Figure 3 shows the photon energy dependence of the near E$_F$ spectra. As previously reported \cite{ShinfirstlaserpaperLa, SatofirstLaHepaper}, the spectra show a sharp discontinuity near the Fermi level which is robust over photon energy.  This discontinuity can be better visualized by symmetrizing the data about E$_F$ to remove the contribution from the Fermi function (Figure 3 inset).  For clarity, it's worth noting that although the symmetrization technique could appear to suggest zero spectral intensity at E$_F$, this is not the case as the main Figure 3 clearly shows.  With the symmetrized data, the discontinuity is manifested by a plateau region, with a width of 30meV (indicated by the blue dotted lines in the inset), which likely represents the previously reported 15meV gap \cite{SatofirstLaHepaper,X.J.Zhoucomparison}.  From our photon energy dependent study, we observe an enhancement of this feature when 50eV photons are used.  This is in agreement with the Fe 3d nature of the near E$_F$ bands, as previously discussed, and is likely due to the vicinity of the photon energy to the Fe 3p resonance at 53eV.   Finally, we note that, given the energy resolution of these data, we do not expect to cleanly resolve the small superconducting gap ($\Delta\sim$4meV at 5K \cite{SatofirstLaHepaper}).

To shed light on the current debate on the origin of the 15 meV gap-like feature \cite{ShinfirstlaserpaperLa, SatofirstLaHepaper, X.J.ZhouFirstLaserSm, X.J.Zhoucomparison}, Figure 4 shows its temperature dependence by performing a high resolution study on the near E$_F$ spectra. The 15meV feature is indicated by the dotted blue arrow in Fig.~4a. Some indications of its temperature dependence are already obvious in the raw data and in the gradual disappearance of the feature near 15 meV in the first derivative spectra (Fig.~4b). To gain more detailed information on the temperature evolution of this feature, we show symmetrized spectra of Figure 4a to once again remove any thermal broadening contribution arising from the Fermi function.  The data show an interesting evolution with temperature, which leads us to identify two characteristics temperatures: 1) T=90K which is the temperature where the gap-like feature "closes up" after a gradual evolution in the line shape with increasing temperature as suggested by the first derivative data (Fig.~4b) and further indicated by the symmetrized data shown in Fig.~4c, 2) A temperature between 110 and 135K where an abrupt change in the shape of the DOS is observed (shown in Fig.~4d), and then remains unchanged for all increasing temperatures, up to at least 180K.

\section{Discussion}
To understand the nature of these two temperatures, we examine the known phase diagram for the LaFeAsO$_{1-x}$F$_{x}$ system, shown in Fig.~4e \cite{LuetkensPhaseLa}.  As discussed in the introduction, the undoped parent compound contains both a magnetic ordering and a structural transition above 100K.  Whether such ordering is suppressed by the Fluorine doping is still controversial as several experimental results seem to suggests its persistence all the way to x=0.1 doping \cite{NowikFirstMossLa,NowikFirstMossSm,AhilanNMRLa,ZhuNernstLa,GonnelliAndreevCoexistence,DrewMuonCoexistence}.  Comparing the extrapolated temperatures for the structural (T$_{STR}$) and SDW transitions (T$_{SDW}$), there is a good agreement between these two temperatures and the two temperatures identified in our data where the DOS undergoes a change.  This leads us to speculate that, although the electron doping of the parent compound clearly competes with the magnetic ordering and structural transition, the coexistence of these phases at higher doping, where superconductivity exists, is in fact occurring and responsible for the observed features.  Thus, we interpret the 15meV feature as a pseudogap-like feature related to the SDW magnetic ordering, which closes around T$_{SDW}$=90K, and we interpret the subtler effect in the DOS around T$_{STR}$=120K as the structural transition seen in the parent compound.

\section{Conclusion}
	   In conclusion, we have used photon energy and temperature dependent angle-integrated photoemission spectroscopy to explore core level peaks, the VB DOS, and near E$_F$ features in polycrystalline LaFeAsO$_{0.9}$F$_{0.1}$.  We have identified the nature of all the core levels peaks and features, and we have provided evidence detailing the orbital character of the VB DOS.  We have found an anomalously large splitting of the As 3d peak and have explained it as a consequence of the anomalously low Fe magnetic moment.  Finally, we have identified two characteristic temperatures: 90K where a pseudogap-like feature closes up and 120K where a more sudden change in the DOS occurs.  We associate these with the SDW magnetic ordering and structural transition temperatures, respectively.  This result suggests that the electron doped superconducting phase does not entirely suppress the magnetic phenomena in the FeAs plane as some recent theory has suggested\cite{NewMazin}.

$^\dagger$ Electronic address: alanzara@lbl.gov

\section{Acknowledgments}
We would like to thank D.-H. Lee, A. Vishwanath, J. Wu, P. Phillips, A. Castro-Neto, S.Y. Zhou, D. Siegel, and O. Fakhouri for helpful discussions. We would also like to thank G. Sawatzky for a critical reading of this manuscript.  The ARPES measurements and data analysis were supported by the Director, Office of Science, Office of Basic Energy Sciences, Materials Sciences and Engineering Division, of the U.S. DOE under Contract No. DE-AC02-05CH11231 and through National Science Foundation through Grant No. DMR03-49361.  Additionally, portions of this research were carried out at the Stanford Synchrotron Radiation Laboratory, a national user facility operated by Stanford University on behalf of the U.S. Department of Energy, Office of Basic Energy Sciences.

\end{document}